\documentclass[pra,notitlepage,twocolumn,superscriptaddress,nofootinbib]{revtex4-1}
\usepackage{amssymb}
\usepackage{bbold}
\usepackage{hyperref}
\hypersetup{plainpages=false,colorlinks=true,linkcolor=blue, citecolor=blue, urlcolor=blue}

\usepackage{pdfpages} 
\makeatletter
\AtBeginDocument{\let\LS@rot\@undefined}
\makeatother 
\usepackage{graphicx}
\usepackage{amsmath}
\usepackage{color}
\usepackage{float} 
\usepackage{multirow}
\usepackage{placeins}
\usepackage{xcolor}
\usepackage{booktabs}
\usepackage{siunitx}
\usepackage{cleveref}
\usepackage{tikz}
\usetikzlibrary{shapes.geometric, arrows}
\pagecolor{white}
\usepackage[braket, qm]{qcircuit}
\tikzstyle{excitOp} = [rectangle, rounded corners, 
minimum width=3cm, 
minimum height=1cm,
text width = 3.5cm,
text centered, 
draw=black, 
fill=blue!25]
\tikzstyle{gateOp} = [rectangle, rounded corners, 
minimum width=2.5cm, 
minimum height=1cm,
text width = 3.5cm,
text centered, 
draw=black, 
fill=blue!25]
\tikzstyle{qcMeasure} = [rectangle, rounded corners, 
minimum height=1cm,
text width = 4cm,
text centered, 
draw=black, 
fill=red!25]
\tikzstyle{noiseFil} = [rectangle, rounded corners, 
minimum height=0.6cm,
text width = 2.3cm,
text centered, 
draw=black, 
fill=green!25]
\tikzstyle{greenFunc} = [rectangle, rounded corners,  
minimum height=1cm,
text width = 5.3cm,
text centered, 
draw=black, 
fill=green!25]
\tikzstyle{arrow} = [thick,->,>=stealth]
\begin{document}
\title{An occupation number quantum subspace expansion approach to compute the single-particle Green function: an opportunity for noise filtering}
\author{B. Gauthier}
\affiliation{D\'epartement de Chimie, Biochimie, Physique et Science Forensique, Institut de Recherche sur l'Hydrog\`ene, Universit\'e du Qu\'ebec \`a Trois-Rivi\`eres, Trois-Rivi\`eres, Qu\'ebec G9A 5H7, Canada}
\author{P. Rosenberg}
\affiliation{D\'epartement de Physique, RQMP \& Institut Quantique, Universit\'e de Sherbrooke, Qu\'ebec, Canada J1K 2R1}
\author{A. Foley}
\affiliation{Institut Quantique, Universit\'e de Sherbrooke, Qu\'ebec, Canada J1K 2R1}
\author{M. Charlebois}
\affiliation{D\'epartement de Chimie, Biochimie, Physique et Science Forensique, Institut de Recherche sur l'Hydrog\`ene, Universit\'e du Qu\'ebec \`a Trois-Rivi\`eres, Trois-Rivi\`eres, Qu\'ebec G9A 5H7, Canada}
\affiliation{D\'epartement de Physique, RQMP \& Institut Quantique, Universit\'e de Sherbrooke, Qu\'ebec, Canada J1K 2R1}
\begin{abstract}
We introduce a hybrid quantum-classical algorithm to compute the Green function for strongly correlated electrons on noisy intermediate-scale quantum (NISQ) devices. The technique consists in the construction of a non-orthogonal excitation basis composed of a set of single-particle excitations generated by occupation number operators. The excited sectors of the Hamiltonian in this basis can then be measured on the quantum device and a classical post-processing procedure yields the Green function in the Lehmann representation. The technique allows for noise filtering, a useful feature for NISQ devices. To validate the approach, we carry out a set of proof-of-principle calculations on the single-band Hubbard model on IBM quantum hardware. For a 2 site system we find good agreement between the results of quantum simulations and the exact result for the local spectral function. This proof-of-principle also shows that the noise filtering provides a reliable way to get rid of satellite peaks present in the spectral weight obtained from a NISQ device. A simulation of a 4 site system carried out on classical hardware suggests that the approach can achieve similar accuracy for larger systems.
\end{abstract}

\maketitle
\section{Introduction}
The Green function is an essential quantity to characterize and understand the behavior of correlated electron systems. Any new method or ground state representation requires a new algorithm to compute the Green function for strongly-correlated electrons. Finding and optimizing these algorithms remain a fundamental and ongoing challenge.
The emergence of quantum computers offers new possibilities for the treatment of strongly-correlated electrons.
These devices hold the promise of more efficient simulations of quantum many-body systems, beyond the
capabilities of classical approaches \cite{Cade2020}. However, decoherence and noise remain significant challenges for
current quantum hardware \cite{Preskill2018}. These limitations prevent the reliable execution of algorithms requiring many 
qubits or gate operations, which makes the development of approaches that can take advantage of these noisy intermediate-scale 
quantum (NISQ) devices an important goal. One class of algorithms designed to run on near-term devices, commonly referred to as 
hybrid algorithms, consist of quantum and classical elements. The most prominent example of such an algorithm being the variational 
quantum eigensolver (VQE) \cite{Peruzzo2014}.     
Here, we introduce a hybrid quantum-classical algorithm to compute the Green function of strongly-correlated electron systems 
on near-term quantum devices. The algorithm uses a non-orthogonal basis expansion technique to measure the excited sectors 
of the Hamiltonian, an approach originally developed within the dynamical variational Monte Carlo (dVMC) framework \cite{Charlebois_2020}, 
which has proven capable of providing accurate results for the Green function of various model systems \cite{Rosenberg2022,Rosenberg2023}.
The technique is an extension of the standard VMC approach based on the mVMC code \cite{misawa_mvmcopen-source_2019}.
The central component of the dVMC technique is the computation of the excited sectors of the Hamiltonian ($N_e\pm1$, where $N_e$ is the number of electrons of the ground state), 
which are used to obtain the Green function in the Lehmann representation. To compute these excited sectors we construct a minimal excitation basis,
composed of states chosen from the set of all possible excited states, i.e. all possible single-particle excitations, according to a physically motivated locality criterion \cite{Rosenberg2022}. This set of excited states is obtained by 
acting a $N_e$ conserving operator on the ground state of the system. 
Although classical dVMC algorithm has good scaling, the choice of variational wave function is arbitrary and potentially biased on very large systems where size renders any benchmark against ED impossible. The need for an alternative and unbiased ground state representation remains. In this article, we replace the classical ground state representation by a quantum circuit to verify if the algorithm still holds with quantum measurement noise (as opposed to classical Monte Carlo noise). We can then evaluate the efficiency of the noise filtering algorithm.
Within the dVMC method, the ground state is obtained via variational Monte Carlo and all matrix elements are computed on classical hardware. 
In the technique we present here, which we will refer to as occupation number quantum subspace expansion (ON-QSE), we employ exact 
representations of the ground state as quantum circuits and compute the necessary matrix elements on quantum hardware, thus making use of Pauli operators acting on quantum circuits in contrast to the manipulation of pair projected BCS states and the computation of Pfaffians for the dVMC method. The remaining 
post-processing components of the algorithm, including a noise-filtering method, are carried out on classical hardware.
The algorithm we present shares features with several algorithms recently introduced to compute the Green function of strongly-correlated electrons
on near-term quantum devices. One set of approaches, based on the idea of variational quantum simulation, carries out real-time evolution on a quantum 
state \cite{Endo_2020,Gomes_2023}, which provides access to the Green function in the real time domain; another, building on the quantum phase estimation technique, 
combines a probabilistic state preparation and measurement technique to compute the frequency dependent Green function \cite{Kosugi_2020}. Other approaches,
referred to as subspace-search methods, compute the excited states of a Hamiltonian by searching a low energy subspace \cite{Nakanishi_2019,Endo_2020,Gyawali_2022}, which can then be used to compute the Green 
function. Another approach is based on the so-called quantum equation of motion \cite{Ollitrault_2020,Rizzo_2022}, which involves
the computation of a set of matrix elements using states generated by excitation operators acting on the ground state. The solution of a generalized 
eigenvalue problem yields the excited states of the Hamiltonian, which can be used to obtain the Green function in the Lehmann representation.
We use exact ground state representation in order to isolate the noise coming from the method we have developed. This is for benchmarking purposes only. While we can also use the variational ground state, we will reserve that for future work. The goal here is to prove the algorithm's potential and its efficiency in filtering noise.
We choose test problems such that the exact ground state representation requires modest circuit depths and numbers of
qubit operations. As we demonstrate below, the new method is fairly robust to noise on the device. The number of matrix elements to be measured
depends on the system size and the number of excitation operators, but for small systems we find that a limited
number of excitation operators are necessary to capture the important features of the spectra. Classical simulations
suggest that the technique can achieve similar results for larger systems, but quantum simulations of these
systems are currently out of reach due to the resources required for such calculations.
We proceed by introducing the method, followed by a set of benchmark calculations on the 2-site Hubbard model
using the $\it{ibm}$\_$\it{sherbrooke}$ quantum computer.

\section{Green function via non-orthogonal basis expansion}
\label{sec:algo}
In the following, we present a hybrid quantum-classical algorithm to compute the single-particle Green function for 
strongly-correlated electrons. 
\subsection{Green function}
The Green function in the Lehmann representation is written:
\begin{align} 
&G_{ij,\sigma}(\omega) = G^{+}_{ij,\sigma}(\omega) + G^{-}_{ij,\sigma}(\omega), \\
&G^{-}_{ij,\sigma}(\omega)=\langle \Omega \vert \hat{c}^\dagger_{i\sigma}\frac{1}{\omega+i\eta-\Omega+\hat{H}}\,\hat{c}_{j\sigma}\vert\Omega\rangle, \\
&G^{+}_{ij,\sigma}(\omega)=\langle \Omega \vert \hat{c}_{i\sigma}\frac{1}{\omega+i\eta+\Omega-\hat{H}}\,\hat{c}^\dagger_{j\sigma}\vert\Omega\rangle,
\end{align}
where $\Omega$ is the energy of the ground state $\ket{\Omega}$, and $\hat{H}$ is the Hamiltonian of the system.
The Green function in this form can be calculated using exact diagonalization (ED), which
can access the complete set of single-particle excited states. Several techniques have been developed 
to compute the ground state and Green function classically for larger systems \cite{Leblanc2015,Schaefer2021}. Here we focus on the dynamical
variational Monte Carlo technique \cite{Charlebois_2020,Rosenberg2022,Rosenberg2023}, which we adapt to take advantage of
near-term quantum devices.
Within the generalized dVMC approach developed in Refs.~\cite{Charlebois_2020,Rosenberg2022,Rosenberg2023}, the Green function matrix at a complex frequency $z$ is computed according to~\cite{Rosenberg2022}:
\begin{align}
\mathbf{G}_{\pm}(z) &= 
\mathbf{S_\pm}((z\pm\Omega)\mathbb{1} \mp\mathbf{H_\pm})^{-1} \mathbf{S_\pm},
\label{eq:Gpm3}
\\
&= \mathbf{Q_\pm}((z\pm\Omega)\mathbb{1} \mp \mathbf{E_\pm})^{-1}\mathbf{Q^\dag_\pm},
\label{eq:Gpm4}
\end{align}
where the matrix $\mathbf{S}_\pm$ is the overlap matrix of the non-orthogonal basis of excited states and $\mathbf{H}_\pm$ is the Hamiltonian matrix. $\mathbf{Q}_\pm\equiv \mathbf{S}_\pm^{1/2}\mathbf{U}_\pm$\footnote{$\mathbf{S}_\pm^{1/2}$ is computed as $\mathbf{S}_\pm^{1/2}=\mathbf{V}_\pm\mathbf{D}_\pm^{1/2}\mathbf{V}_\pm^\dagger$, where $\mathbf{V}_\pm$ is a matrix of the eigenvectors of $\mathbf{S}_\pm$, and $\mathbf{D}_\pm$ is the diagonal matrix of eigenvalues.}  and $\mathbf{U}_\pm$ and $\mathbf{E}_\pm$ are the eigenvectors and eigenvalues
respectively of the matrix $\mathbf{M}_\pm \equiv \mathbf{S}_\pm^{-1/2}\mathbf{H}_\pm\mathbf{S}_\pm^{-1/2}$. The matrix elements of $\mathbf{S}_\pm$  and  $\mathbf{H}_\pm$ are defined as:
\begin{align}
S^+_{im\sigma,jn\sigma^\prime} &= \bra{\psi_{im\sigma}} \hat{c}_{i\sigma}\hat{c}^\dagger_{j\sigma^\prime} \ket{\psi_{jn\sigma^\prime}}, \label{eq:Se_k}\\ 
S^-_{im\sigma,jn\sigma^\prime} &= \bra{\psi_{im\sigma}} \hat{c}^\dagger_{i\sigma}\hat{c}_{j\sigma^\prime} \ket{\psi_{jn\sigma^\prime}},\label{eq:Sh_k}
\end{align}
and,
\begin{align}
H^+_{im\sigma,jn\sigma^\prime} &= \bra{\psi_{im\sigma}} \hat{c}_{i\sigma}\hat{H} \hat{c}^\dagger_{j\sigma^\prime}\ket{\psi_{jn\sigma^\prime}}, \label{eq:He_k} \\
H^-_{im\sigma,jn\sigma^\prime} &= \bra{\psi_{im\sigma}} \hat{c}^\dagger_{i\sigma}\hat{H} \hat{c}_{j\sigma^\prime}\ket{\psi_{jn\sigma^\prime}}, \label{eq:Hh_k}
\end{align}
where the state $\ket{\psi_{im\sigma}}\equiv\hat{B}_{im\sigma}\ket{\Omega}$ is formed by acting an excitation operator, $\hat{B}_{im\sigma}$, on the many-body ground state, 
$\ket{\Omega}$. The operator $\hat{B}_{im\sigma}$ can be any $N_e$ conserving operator. This set of excitation operators generates a non-orthogonal basis of excited states, which is used to calculate the Green function according to Eq.~(\ref{eq:Gpm4}). The operators are chosen to span a physically important subspace of the full single-particle excitation basis. Previous experience \cite{Charlebois_2020,Rosenberg2022} has indicated that taking combinations of number operators for the $\hat{B}_{im\sigma}$ can produce an easy to compute restricted excitation basis that includes highly relevant states. 
In the ON-QSE method, we choose the operators,
\begin{eqnarray}
&&\hat{B}_{i0\sigma} = \mathbb{1} 
\label{eq:exc_states1}
\\
&&\hat{B}_{i1\sigma} = \hat{n}_{i\bar{\sigma}} 
\label{eq:exc_states2}
\\
&&\hat{B}_{im\sigma} = \hat{n}_{b_{im},\bar{\sigma}}\hat{n}_{b^\prime_{im},\sigma}, \quad\quad\quad m \geq 2,
\label{eq:exc_states}
\end{eqnarray}
where $b_{im}$ and $b_{im}^\prime$ are a list of different labels of the neighboring sites to $i$, as illustrated in Ref~\cite{Charlebois_2020,Rosenberg2022}. As discussed in these references, the specific choice of $(b_{im}, b_{im}^\prime)$  does not matter too much, due to the redundant information captured by the overlapping basis; a different choice of $(b_{im}, b_{im}^\prime)$ will lead to a very similar result. Fig. 8 of Ref.~\cite{Charlebois_2020} studies the
convergence with respect to the number of excitations in the context of dVMC and shows that there is a saturation when using more and more excitations. Here, we choose a sufficient number of excitations, such that the result does not change meaningfully with the inclusion of additional excitations.
This saturation is possible because the basis is non-orthogonal, and a big part of the information is shared across multiple states. 
There is no systematic way to predict if the number of excitations is sufficient. However, we know that the number of peaks in the spectra computed using our formalism will at most be equal to the number of excitations used. The general idea is to select a number of excitations greater than the number of peaks we expect to appear in the spectrum,  and then verify in post processing that the resulting local density of states is stable by manually removing one or two excitations. It is also useful to select more excitations in order to improve the filtering algorithm described below. 

The Green function is obtained from the Green function matrix (Eq.~(\ref{eq:Gpm4})) according to:
\begin{align}
G_{ij,\sigma}(z) = \left[ \mathbf{G}_+(z) + \mathbf{G}_-(z) \right]_{ij,\sigma=\sigma^\prime,m=n=0},
\label{eq:G_ij}
\end{align}
 by virtue of the fact that we take $m=0$ to be the trivial excitation (i.e. $\ket{\psi_{i0\sigma}}=\mathbb{1}\ket{\Omega}$).
 
The matrices $\mathbf{H}_\pm$ and  $\mathbf{S}_\pm$ have dimensions $NN_{exc}\times NN_{exc}$, meaning that there are $\mathcal{O}(N^2N_{exc}^2)$ terms to be computed on the quantum device. In order to reduce the required computational resources, we use only few different excitations $\hat{B}_{im\sigma}$. In general, the number of excitations can be reduced by exploiting symmetries of the system.
\subsection{Filtering method}
\begin{figure}[!ht]
\begin{center}
\includegraphics[width=0.95\columnwidth]{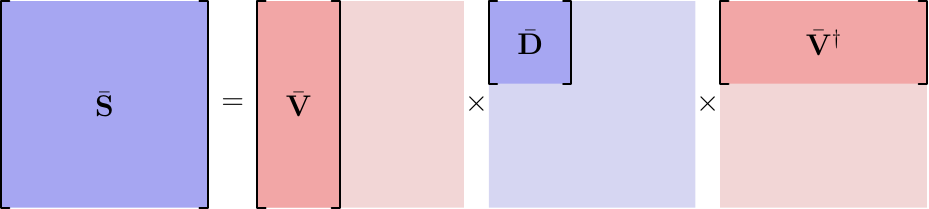}
\end{center}
\caption{Illustration of filtering procedure. The pale red and blue squares show the original ${\bf V}$ and ${\bf D}$
matrices, respectively. The matrix $\bar{\bf D}$ contains only positive eigenvalues. Note that the matrix $\bar{\bf S}$
has the same dimension as the original matrix ${\bf S}$.}
\label{fig:filtering}
\end{figure}

In order to mitigate the effects of quantum noise in the measurement of the matrix elements, we carry out a classical noise filtering procedure \cite{Rosenberg2022} during the 
post-processing phase of the algorithm. Here, we omit the $\pm$ for the sake of brevity. Once the matrix elements of $\mathbf{S}$ have been measured, we compute its eigendecomposition, $\mathbf{S} = \mathbf{V} \mathbf{D} \mathbf{V}^\dag$, where $\mathbf{D}$ is a diagonal matrix of eigenvalues and the matrix $\mathbf{V}$ contains the corresponding eigenvectors. We then truncate $\mathbf{D}$ so that it contains only positive eigenvalues, and similarly truncate the corresponding columns of $\mathbf{V}$ and rows of $\mathbf{V}^\dagger$. Finally, we construct the filtered matrix $\mathbf{\bar S} = \mathbf{\bar V} \mathbf{\bar D} \mathbf{\bar V}^\dag$, as a product of the truncated matrices, $\mathbf{\bar V}$, $\mathbf{\bar D}$, and $\mathbf{\bar V}^\dagger$. Note that $\mathbf{S}$ and $\mathbf{\bar S}$ are of the same dimension (see Fig.~\ref{fig:filtering}). The filtered matrix $\mathbf{\bar S}$ then replaces $\mathbf{S}$ in Eq.(\ref{eq:Gpm3}), which affects the resulting matrices $\mathbf{M}$, $\mathbf{U}$, $\mathbf{E}$ and $\mathbf{Q}$. It is to be noted that this filtering algorithm can be applied to any Green function calculation algorithm involving the solution of a noisy generalized eigenvalue problem.
The filtering method works on the principle that any overlap matrix $\mathbf{S}$ of a non-orthogonal basis, as defined in Eqs.~\eqref{eq:Se_k} and~\eqref{eq:Sh_k}, in the absence of noise, is positive definite. However, the eigenvalues of this matrix can become very small. In the presence of noise, these small eigenvalues can even become negative. The computation of $\mathbf{M}_\pm$ requires the matrix $\mathbf{S}_\pm^{-1/2}$, which cannot be computed when $\mathbf{S}_\pm$ has negative eigenvalues, and can diverge when the eigenvalues are too small, which effectively increases the polluting effect of the Monte Carlo sampling noise or quantum noise. The filtering algorithm works by identifying which combination of overlapping states needs to be omitted in order to get rid of the small or negative eigenvalues. We denote the tolerance threshold on the eigenvalues of $\mathbf{S}_\pm$ as $\alpha$, and illustrate its role in Sec. \ref{sec:N2mu2}. Note that this filtering method also removes any redundant or equivalent states, because they would produce a null eigenvalue.
As alluded to above, several related methods to compute the Green function have recently been introduced.
In particular, the quantum equation of motion approach \cite{Ollitrault_2020,Rizzo_2022} is built upon a similar
concept to the technique we have developed, specifically the computation of excited states using a limited set
of excitation operators. However, unlike the quantum equation of motion approach \cite{Rizzo_2022}, we include only occupation number
operators in the set of excitation operators, which translates into a more straightforward calculation of the Green function as we don't need to compute double commutators as in Ref.~\cite{Ollitrault_2020,Rizzo_2022}. Additionally, we find that computing the Green function via Eq.~(\ref{eq:Gpm4}) and
applying a noise filtering procedure yields an excitation spectrum that is relatively robust to noise in the measurements of
the matrix elements, Eqs.~(\ref{eq:Se_k})-(\ref{eq:Sh_k}), (\ref{eq:He_k})-(\ref{eq:Hh_k}). Compared to the equation of motion approach Ref.~\cite{Ollitrault_2020,Rizzo_2022}, ON-QSE is free from the bias of a variational ansatz; it doesn't require devising an ansatz
that is able to represent excited states as well as the ground state. 
\subsection{Ground state representation}
\begin{figure}
    \begin{center}
        \begin{tikzpicture}[node distance=1.8cm]
        
        \node (1st) [gateOp] {Gate operations to construct the exact ground state $\ket{\Omega}$};
        \node (2nd) [excitOp, right of=1st, xshift=2.5cm] {Classical calculation of $\hat{B}_{im\sigma}^\dag\hat{c}_{i\sigma}\hat{H} \hat{c}^\dagger_{j\sigma^\prime}\hat{B}_{jn\sigma^\prime}$};
        \node (3rd) [qcMeasure, below of=1st, xshift=2cm, yshift=0.1cm] {Measuring $\mathbf{S}_\pm$ and $\mathbf{H}_\pm$ on the quantum computer};
        \node (4th) [noiseFil, below of=3rd, yshift=0.5cm] {Noise filtering};
        \node (5th) [greenFunc, below of=4th, yshift=0.5cm] {Green function calculation using filtered $\mathbf{S}_\pm$ and $\mathbf{H}_\pm$ matrices};
        
        \draw [arrow,ultra thick] (1st) -- (3rd);
        \draw [arrow,ultra thick] (2nd) -- (3rd);
        \draw [arrow,ultra thick] (3rd) -- (4th);
        \draw [arrow,ultra thick] (4th) -- (5th);
        
        \end{tikzpicture}
    \end{center}
    \caption{Flowchart expressing the steps involved in the ON-QSE algorithm. The matrix elements $S^+_{im\sigma,jn\sigma^\prime}$, $S^-_{im\sigma,jn\sigma^\prime}$, $H^+_{im\sigma,jn\sigma^\prime}$ and $H^-_{im\sigma,jn\sigma^\prime}$ are defined by Eqs. (\ref{eq:Se_k})-(\ref{eq:Sh_k}), (\ref{eq:He_k})-(\ref{eq:Hh_k}) and $\hat{B}_{im\sigma}^\dag$ can be any $N_e$ conserving operator. $\hat{H}$ is the Hamiltonian, $\hat{c}_{i\sigma}^\dag$ the creation operator and $\hat{c}_{i\sigma}$ the annihilation operator. Classical pre-processing, quantum computer processing and classical post-processing are respectively colored in blue, red and green.}
    \label{fig:flowchart}
\end{figure}
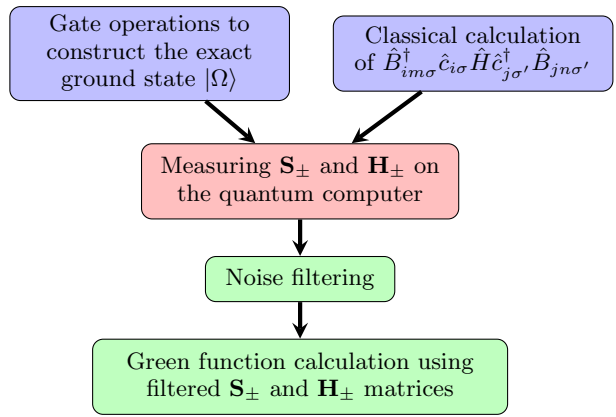

As stated above, this exact ground state representation is used for benchmark purposes only, as we expect the number of gates of the exact ground state to scale exponentially with system size. However, it provides an unbiased approach to estimate the quantum noise coming from our method, and the efficiency of the noise filtering algorithm.
We emphasize that the algorithm is quite flexible in terms of both the choice of excitation operators and the form of the ground state. In the dVMC method, the ground state is obtained via variational Monte Carlo, and the matrix elements in Eqs.~(\ref{eq:Se_k})-(\ref{eq:Sh_k}) and (\ref{eq:He_k})-(\ref{eq:Hh_k}) are estimated using a Metropolis-based Markov chain process. The filtering approach we have applied is also an important innovation that can provide considerably improved results from measurements on noisy quantum devices. While this approach has proven capable of treating Monte Carlo noise, it was not known before implementing and testing this algorithm on a quantum device that this procedure would be effective for filtering quantum device noise.
In the present work we choose to represent the ground state exactly as a quantum circuit (see Figs.~\ref{fig:2sites_mu0}(a),\ref{fig:2sites_mu2}(a),\ref{fig:4sites_mu0}(a)) and use the exact ground state energy in Eqs.~(\ref{eq:Gpm3})-(\ref{eq:Gpm4}), in order to better gauge the accuracy of the method without introducing additional error from the ground state computation, but the same method can be applied to a ground state obtained via the VQE, for instance. The operators in Eqs.~(\ref{eq:Se_k})-(\ref{eq:Sh_k}) and (\ref{eq:He_k})-(\ref{eq:Hh_k}) are then mapped to strings of Pauli operators via the Jordan-Wigner transformation \cite{JordanWigner}, and the corresponding matrix elements are measured on the quantum device. The code was implemented using the open-source Python 3 library, Qiskit \cite{Qiskit}, which can simulate and launch programs on IBM quantum devices. The remaining steps of the algorithm are executed on classical hardware. Fig. \ref{fig:flowchart} expresses the principal steps involved in our algorithm.
For later convenience, we introduce the following notation for qubit operations:
\begin{align}
 (X_i) =& \ket{0_i}\bra{1_i}+\ket{1_i}\bra{0_i}\\
 (C_iX_j) =& \ket{0_i}\bra{0_i} \mathbb{1}_j + \ket{1_i}\bra{1_i} X_j\\
 (H_i) =& \ket{+_i}\bra{0_i} + \ket{-_i}\bra{1_i} \\
 \ket{\pm_i} =& \frac{1}{\sqrt{2}}\left[ \ket{0_i} \pm \ket{1_i}\right] 
\end{align}
$(X_i)$ represents the bitflip operator acting on the $i$th qubit, corresponding to the Pauli matrix $\sigma_x$. $(C_iX_j)$ is the controlled bitflip operator, where the $i$th qubit is the control and the $j$th qubit is the target. $(H_i)$ is the Hadamard gate acting on the $i$th qubit. Additional operator definitions are provided in Appendix~\ref{app:GS_N2mu2}.

\section{Model}
To study the performance of the technique we carry out calculations on the single-band Hubbard model:
\begin{equation}
\hat{H} = -t\sum_{\langle ij \rangle,\sigma} ( \hat{c}^\dagger_{i\sigma}\hat{c}_{j\sigma} + \text{h.c.}) + U\sum_i \hat{n}_{i\uparrow} \hat{n}_{i\downarrow} + \mu\sum_i\hat{n}_{i\sigma},
\end{equation}
where $\hat{c}^\dagger_{i\sigma}$ creates an electron of spin $\sigma$ at site $i$ and $\hat{n}_{i\sigma}\equiv \hat{c}^\dagger_{i\sigma}\hat{c}_{i\sigma}$. We take the nearest neighbor hopping $t=1.0$ and the on-site Hubbard interaction strength $U=4.0$.
In the following, we present results for 2 and 4 site Hubbard chains.
We focus on the local spectral function $A_{ii,\sigma}(\omega) = -(1/\pi)\textrm{Im}[G_{ii,\sigma}(\omega)]$.

\section{Results}
All quantum simulations were performed on the $\it{ibm}$\_$\it{sherbrooke}$ quantum device using the Qiskit Runtime Estimator with default resilience and optimization levels. Calibration data for the results are presented in Appendix~\ref{app:calibration}.
\subsection{$N=2$, $\mu=0$}
\begin{figure}[!ht]
\begin{center}
\includegraphics[width=0.95\columnwidth]{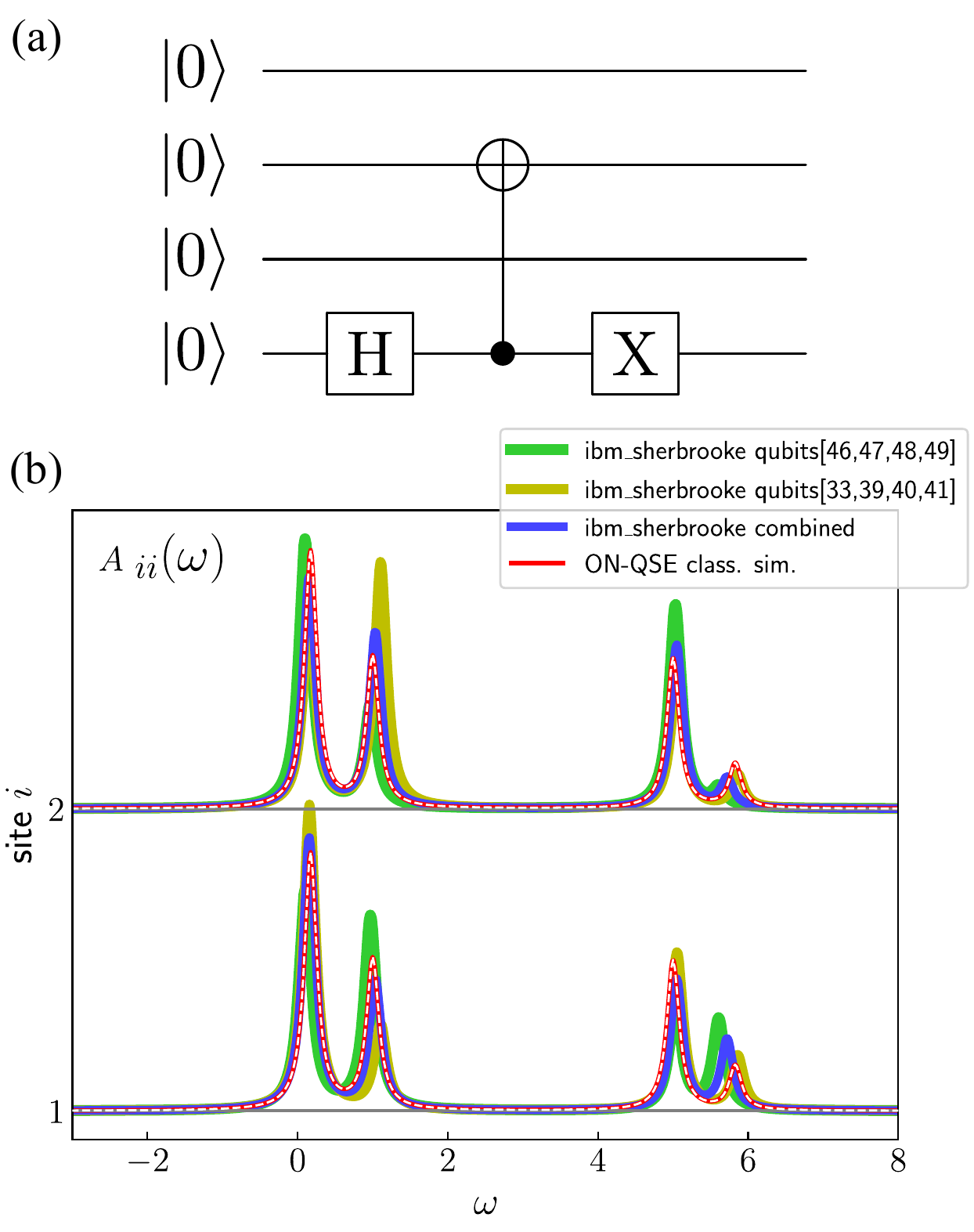}
\end{center}
\caption{Circuit diagram (a) and local spectral function (b) for the ground state of the 2 site Hubbard model at $\mu = 0.0$.
The excitation operators used were $\hat{B}_{1m\downarrow} = \{\mathbb{1}$, $\hat{n}_{1\uparrow}$\} and  $\hat{B}_{2m\downarrow} = \{\mathbb{1}$, $\hat{n}_{2\uparrow}$\}. The dashed white curve corresponds to the ED result.
}
\label{fig:2sites_mu0}
\end{figure}

We begin with the case of $\mu = 0$.
In this case, the ground state is represented by the quantum circuit illustrated in Fig.~\ref{fig:2sites_mu0}(a). This circuit representation is obtained by transforming the exact ground state written in the Fock basis.
At $\mu = 0$, the ground state belongs to the $N_e = 1$, $S_z=\pm 1$ (degenerate) particle sectors, as determined by ED calculations. Here we choose $S_z = 1$.
In the basis $\ket{2\hspace{-0.1cm}\uparrow,2\hspace{-0.1cm}\downarrow,1\hspace{-0.1cm}\uparrow,1\hspace{-0.1cm}\downarrow}$ the ground state for this system is:
\begin{equation}
\frac{1}{\sqrt{2}}\left(\ket{1000}+ \ket{0010}\right).
\end{equation}
To transform this state from the Fock basis representation to the circuit representation shown in Fig.~\ref{fig:2sites_mu0}(a)
we perform a series of one and two qubit operations (Note that we order qubits from right to left). Explicitly,
 \begin{align}
\frac{1}{\sqrt{2}}\left(\ket{1000}+ \ket{0010}\right) &= \frac{1}{\sqrt{2}}(X_3)\left(\ket{0000}+\ket{1010}\right) \notag \\
&= \frac{1}{\sqrt{2}}(X_3)(C_3 X_1)\left(\ket{0000}+\ket{1000}\right) \notag \\
&= (X_3)(C_3 X_1)(H_3)\ket{0000}.
\end{align}
In Fig.~\ref{fig:2sites_mu0}(b) we show the local spectral function for spin down, $A_{ii,\downarrow}(\omega)$.
The blue curve shows the result from a calculation run on the $\it{ibm}$\_$\it{sherbrooke}$ quantum device.
The red curve shows the result of a classical simulation using the ON-QSE technique, starting from the 
exact ground state written in the Fock basis. The dashed white curve corresponds to the ED result.
We first examine the classical simulation of the technique (red curve), which uses the 
exact ground state written in the Fock basis. The result shows perfect agreement with the
ED result, indicating that the set of excitations is sufficient to reproduce the expected physics,
and that any deviations from the ED result appearing in the quantum simulation are related to
noise from the quantum device. Now considering the results of the quantum simulation (blue curve),
we find good agreement with the exact result. The structure and spacing of the peaks is reliably captured, although there are small displacements of the peaks and also some secondary noise peaks. These irregularities in the spectrum are explained by the small system size, which means that the number of excitation operators acting on the ground state is also small. This gives less flexibility to the noise filtering algorithm to filter out noisy states. In order to get rid of the quantum noise, the measurements were repeated twice on two different sets of qubits, and the matrices $\mathbf{S}_\pm$ and $\mathbf{H}_\pm$ from these two measurements were averaged in post processing. 
As we can see from Fig. \ref{fig:2sites_mu0}, the results produced by each individual set are noticeably less accurate than their average. Thus, in the absence of noise filtering, it was necessary to combine different sets of measurements in order to reduce the quantum noise. The next section shows that this combination is not necessary when we use noise filtering instead.
\subsection{$N=2$, $\mu=2$} \label{sec:N2mu2}
We proceed with a calculation of the local spectral function for spin up, $A_{ii,\uparrow}(\omega)$, for
$\mu = 2$, corresponding to the $N_e=2$, $S_z=0$ particle sector. The quantum circuit chosen to represent the
ground state in this case is illustrated in Fig.~\ref{fig:2sites_mu2}(a) and the details of its calculation can be found in Appendix~\ref{app:GS_N2mu2}. When the tolerance threshold $\alpha = 10^{-8}$, the filtering is only enforcing that the matrices $\mathbf{S}_\pm$ are positive definite and the effect of the filtering is minimal. Thus the spectrum deviates significantly from the ED result. 
However, with $\alpha = 0.05$, 
the spectrum agrees with ED. 
We chose the value of $\alpha$ by looking at the standard deviation of the measured $\mathbf{S}_\pm$ matrices, which was around $0.02$. The result from this filtering step is striking, both removing satellite peaks always present in the quantum computer spectral weight calculation and correcting for split peaks.
\begin{figure}[!ht]
\begin{center}
\includegraphics[width=0.95\columnwidth]{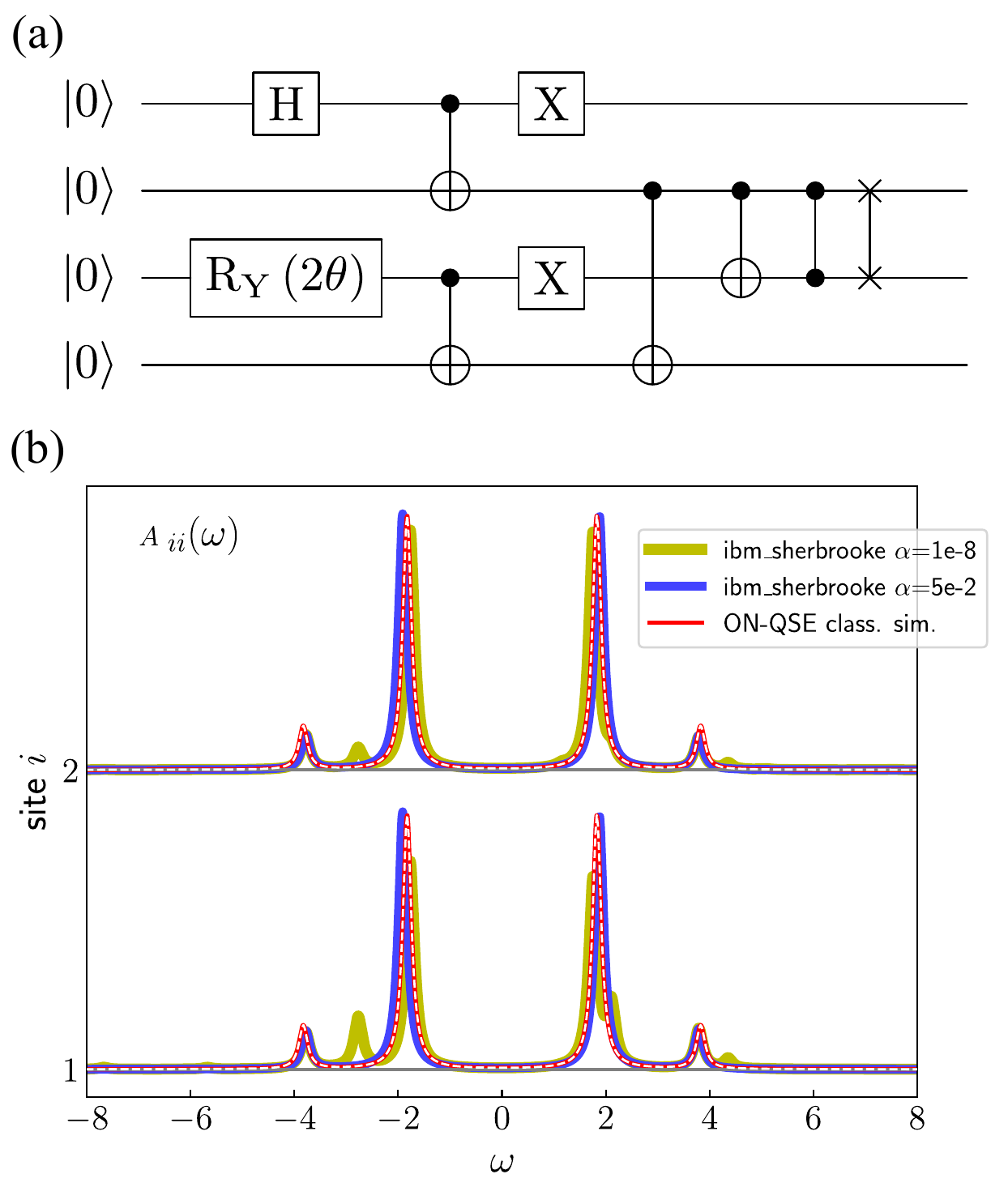}
\end{center}
\caption{Circuit diagram (a) and local spectral function (b) for the ground state of the 2 site Hubbard model at $\mu = 2.0$. The excitation operators used were $\hat{B}_{1m\uparrow} = \{\mathbb{1}$, $\hat{n}_{1\downarrow}$, $\hat{n}_{1\downarrow}\hat{n}_{2\uparrow}$, $\hat{n}_{2\downarrow}\hat{n}_{2\uparrow}$ \} and  $\hat{B}_{2m\uparrow} = \{\mathbb{1}$, $\hat{n}_{2\downarrow}$, $\hat{n}_{2\downarrow}\hat{n}_{1\uparrow}$, $\hat{n}_{1\downarrow}\hat{n}_{1\uparrow}$\}. The operators of the type $\hat{n}_{i\downarrow}\hat{n}_{j\uparrow}$ are now possible here compared to Fig.~\ref{fig:2sites_mu0}, since $\hat{n}_{i\downarrow}\hat{n}_{j\uparrow}\ket{\Omega}$ is finite, contrary to the single electron case when $\mu = 0$. The dashed white curve corresponds to the ED result.
}
\label{fig:2sites_mu2}
\end{figure}
\subsection{$N=4$, $\mu=0$}
\label{sec:N4mu0}
The memory restrictions of current quantum hardware prevent us from carrying out quantum simulations on larger systems, however, the hybrid quantum-classical algorithm requires a similar number of operations to the fully classical version, which
has polynomial scaling with the number of sites and excitations \cite{Rosenberg2022},
suggesting
that improvements to the memory 
capacity of quantum devices, or a more memory-efficient implementation of the algorithm, should enable simulations of larger systems.
As an illustration of the potential of the method, we conclude with a classical simulation of the quantum algorithm for a 4 site system at 
$\mu = 0.0$, corresponding to the $N_e=2$, $S_z=0$ particle sector. The circuit representing the ground state for this system is shown in Fig.~\ref{fig:4sites_mu0} (see Appendix~\ref{app:GS_N4mu0} for the origin of the circuit). 
We observe perfect agreement between the ED result (dashed white curve)
and the classical simulation starting from a ground state in the Fock basis (red curve), as well as the classical simulation of
the quantum algorithm, again suggesting that the set of excitations we have chosen is adequate. We therefore expect
any noise in a quantum simulation to be a result of device noise, which might be reduced somewhat by hardware improvements and more advanced error mitigation (see Appendix~\ref{app:results_N4mu0} for 4 site results obtained from the $\it{ibm}$\_$\it{sherbrooke}$ device, which provide an indication of the current level of noise present in larger simulations).
\begin{figure}[!ht]
\begin{center}
\includegraphics[width=0.975\columnwidth]{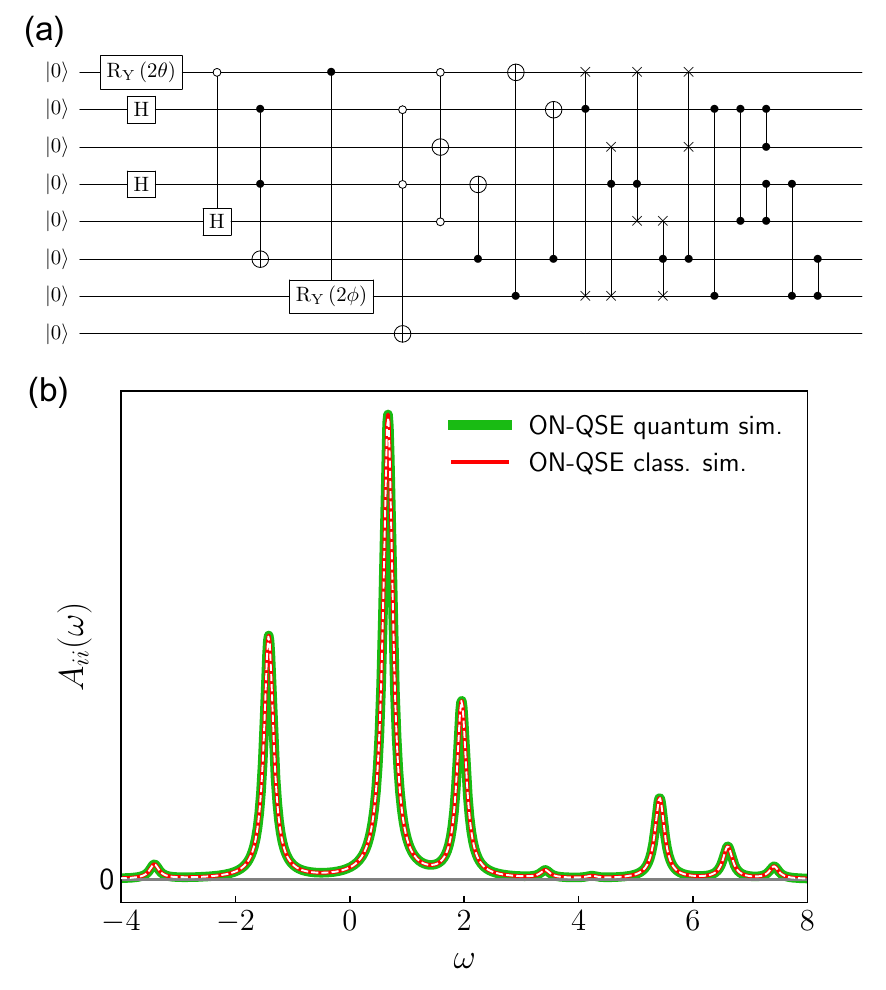}
\end{center}
\caption{Circuit diagram (a) and local spectral function (b) for the ground state of the 4 site Hubbard model at $\mu = 0.0$.
For site 1, the set of excitation operators used is $\hat{B}_{1m\uparrow} = \{\mathbb{1}$, $\hat{n}_{1\downarrow}$, $\hat{n}_{1\downarrow}\hat{n}_{2\uparrow}$,
$\hat{n}_{2\downarrow}\hat{n}_{2\uparrow}$, $\hat{n}_{2\downarrow}\hat{n}_{3\uparrow}$, $\hat{n}_{3\downarrow}\hat{n}_{2\uparrow}$\}.
An equivalent set of excitations, related by symmetry, is used for the remaining sites.
The dashed white curve corresponds to the ED result.
}
\label{fig:4sites_mu0}
\end{figure}

\subsection{Scaling}
The part of the quantum circuit necessary to measure the Green function has constant depth, regardless of the form of the ground state (exact or variational). This is due to the fact that the different excitations operators $\hat{B}_{im\sigma}$ are only, at most, the product of two number operators. 
The circuit depth does not change with the size of the system $N$, and the number of excitations chosen, $N_{exc}$. However, the number of circuits (values) to measure is proportional to $N^2 \times N_{exc}^2$. The bottleneck of this algorithm is really the number of these values that one can measure, which saturates quite quickly on present quantum computers, as discussed in Sec.~\ref{sec:N4mu0}. Also, with more memory will come the possibility to add more excitations in order to increase the performance of the noise filtering feature of the algorithm.
The natural next step is to implement this algorithm with a variational ground state quantum circuit. This will introduce another source of noise to the calculation, but the depth of the circuit will be more manageable.
\subsection{Filtering Advantage}
Our results clearly illustrate the role and importance of the filtering method we have implemented. Fig.~\ref{fig:2sites_mu2} reveals that this technique is a robust solution capable of removing satellite peaks in the density of states caused by quantum noise. It is fairly easy to estimate the noise level from the sampling method of any NISQ device.  
To our knowledge, this is the first proposed algorithm to allow for the filtering of quantum device noise in the computation of the Green function. As discussed in Sec. \ref{sec:algo}, the filtering procedure is designed to remove the redundant information contained in the overlapping basis, thereby minimizing the noise in the computed Green function. It is not clear that this procedure can be directly applied in other approaches to compute the Green function, except for those that use a non-orthogonal basis, as in Ref.~\cite{Rizzo_2022}, which seems to suggest the usefulness of such a basis in future methods.
\section{Conclusion}
We have introduced a hybrid quantum-classical algorithm to compute the Green function for strongly-correlated 
electrons that can be run on current and near term quantum devices. The method is based on the construction of a non-orthogonal 
basis of excited states, generated by acting number operators on the ground state. The excited sectors of the Hamiltonian in this basis 
are then measured on the quantum device, followed by a post-processing procedure carried out on the classical device that provides
the Green function in the Lehmann representation. As a proof of principle, we compute the local spectral function for the
single-band Hubbard model. We find that results from simulations run on the quantum device are in good agreement with
the exact result for 2 site systems. Classical simulations for a 4 site system indicate that the approach should achieve
similar accuracy for larger systems, which are currently out of reach due to memory limitations.
The advent of quantum computers has the potential to open new avenues in the field of strongly correlated systems.
While many challenges limit the current applications of quantum hardware, the development of methods that are
robust to noise is an essential goal. The technique we have developed here serves as both a demonstration of
the possibilities for current devices, as well as a suggestion of what might be achievable with improved algorithms
and hardware. We envision several pathways for future work, including the implementation of more advanced quantum error
mitigation strategies \cite{cai_quantum_2023}, as well as more efficient approaches to performing measurements \cite{nakaji_measurement_2023,zhang_composite_2023,yen_cartan_2021,choi_measurement_2023,choi_improving_2022}, aiming to reduce both
the number and depth of quantum circuits required by the algorithm. 
Additionally, ON-QSE's effectiveness has been rigorously tested using exact ground states. Now, we are considering evaluating the resilience of our method against approximate ground states, which would provide us with more insights into the scalability of the technique relative to the size of the simulated lattice. Notably, one way to proceed would be to prepare quantum circuit representations of the ground states using a quantum variational circuit. 
\begin{acknowledgments}
B.G. was supported by the Natural Sciences and Engineering Research Council (Canada) under Grant Nos. RGPIN-2021-04043. 
He wrote most of the code using the Qiskit library and performed most of the initial calculations.
P.R. was supported by a postdoctoral fellowship from the Canada First Research Excellence Fund through Institut quantique. He finalized the calculations initiated by B.G. and wrote most of the article.
We thank the Quantum AlgoLab at Institut Quantique for providing quantum computing resources and technical support.
\end{acknowledgments}
\appendix
\section{Qubit calibration}
\label{app:calibration}

\subsection{$N=2$, $\mu=0$}
The measurement of the $N=2$, $\mu=0$ (sector $N_e=1$, $S_z=1$) Green function was performed with two different sets of qubits, on two different days. We did 10,000 measurements per observable on qubits $\{46,47,48,49\}$ and another 10,000 measurements per observable on qubits $\{33,39,40,41\}$. The result presented in Fig.~\ref{fig:2sites_mu0} represents the total average of 20,000 measurements. 
The
result is better after averaging $\mathbf{S}_\pm$ and  $\mathbf{H}_\pm$ measured using the two different sets of qubits, suggesting that averaging over different qubit configurations cancels some of the bias present in a given qubit configuration. This is necessary only if one uses too few excitations to take full advantage of the noise filtering algorithm.
%
One can find the relevant qubits and gates average properties and standard deviation in \cref{tab:Gatp_sesA,tab:Gatp_sesB,tab:qbitp_sesA,tab:qbitp_sesB}. On each of these days, the quantum computer's calibration was updated frequently. This is explicited in the table as an uncertainty to each calibration data.
\begin{table}[ht!]
\caption{Qubits properties, single electron simulation, qubit set A}
\label{tab:qbitp_sesA}
\begin{tabular}{cllll}
\hline
\hline
 qubits & T1 (\si{\micro \second}) & T2 (\si{\micro \second}) &  Freq. (\si{\giga \hertz}) &  anharm. (\si{\giga \hertz})  \\
\hline
$46$ & $405\pm61$ & $306$   & $4.674059 (9)$ & $-0.3133505$ \\
$47$ & $444\pm30$ & $185\pm14$ & $4.7950598 (21)$ & $-0.3122514$ \\
$48$ & $289\pm44$ & $278\pm5$  & $4.7072340 (7)$ & $-0.3116499$ \\
$49$ & $239\pm10$ & $145\pm3$  & $4.8107556 (5)$ & $-0.3111314$ \\
\hline
\hline
\end{tabular}
\end{table}
\begin{table}[ht!]
\caption{Gate properties, single electron simulation, qubit set A}
\label{tab:Gatp_sesA}
\begin{tabular}{llll}
\hline
\hline
 qubit & ECR error & Readout error & $1$qb gate error \\
\hline
$46$ &$46\leftrightarrow47 : 0.0022(3) $ &$0.009(4)$  & $0.000091(5)$ \\
$47$ &$47\leftrightarrow48 : 0.0033(5) $ &$0.0103(4)$ & $0.00011(3)$ \\
$48$ &$48\leftrightarrow49 : 0.0087(9) $ &$0.0108(2)$ & $0.000153(9)$ \\
$49$ &                                   &$0.0072(5)$ & $0.00025(6)$ \\
\hline
\hline
\end{tabular}
\end{table}

\begin{table}[ht!]
\caption{Qubits properties, single electron simulation, qubit set B}
\label{tab:qbitp_sesB}
\begin{tabular}{lllll}
\hline
\hline
 qubits & T1 (\si{\micro \second}) & T2 (\si{\micro \second}) &  Freq. (\si{\giga \hertz}) &  anharm. (\si{\giga \hertz})  \\
\hline
$33$ & $510\pm 41$ & $233$       & $4.6775110(2)$ & $-0.3137282$ \\
$39$ & $449\pm165$  & $178\pm26$ & $4.574200(8)$ & $-0.3140308$ \\
$40$ & $410\pm 30$  & $337\pm17$ & $4.7055210(7)$ & $-0.3126427$ \\
$41$ & $209\pm 18$  & $278\pm60$ & $4.8145392(8)$ & $-0.3126369$ \\
\hline
\hline
\end{tabular}
\end{table}

\begin{table}[ht!]
\caption{Gate properties, single electron simulation, qubit set B}
\label{tab:Gatp_sesB}
\begin{tabular}{llll}
\hline
\hline
 qubit & ECR error & Readout error & $1$qb gate error \\
\hline
33 &$33\leftrightarrow39: 0.0046(9)$  &$0.0055(6)$  & $0.00013(12)$ \\
39 &$39\leftrightarrow40: 0.0046(22)$ &$0.008(2)$ & $0.00022(8)$ \\
40 &$40\leftrightarrow41: 0.0070(16)$ &$0.0045(1)$  & $0.00012(4)$ \\
41 &$ $                               &$0.0047(4)$  & $0.00018(4)$ \\
\hline
\hline
\end{tabular}
\end{table}

\subsection{$N=2$, $\mu=2$}
The measurement of the $N=2$, $\mu=2$ (sector $N_e=2$, $S_z=0$) Green function was performed with only a single sets of qubits  $\{60,59,58,71\}$  with 10,000 measurements per observable.
The relevant properties of the quantum computer at the time of these computations can be found in \cref{tab:qbitp_tes,tab:Gatp_tes}. Note that during the simulation, the reported T1 of qubit 59 and 71 increased to \SI{438.407}{\micro\second} and \SI{448.969}{\micro\second}, respectively. The other properties, as reported by IBM quantum, did not change for the duration of the computations, thus the tables contain no uncertainties.
These latter figures likely reflect more accurately the condition of the simulation, since they are more recent.
Because we estimate our circuit's duration to be less than \SI{10}{\micro \second}, the effect of this change of each of the qubits' contribution to the error would be of the order of $1\%$, far less than other error sources.
\begin{table}[ht!]
\centering
\caption{Qubit properties, two electrons simulation\label{tab:qbitp_tes}}
\begin{tabular}{lllll}
\hline
\hline
qubit & T1 (\si{\micro \second}) & T2 (\si{\micro \second}) &  Freq. (\si{\giga \hertz}) &  anharm. (\si{\giga \hertz})  \\
\hline
$ 60  $&$  281.44  $&$  186.19 $&$ 4.672353 $&$ -0.313240 $ \\
$59   $&$  264.90  $&$  445.56 $&$ 4.810886 $&$ -0.310236 $ \\
$58   $&$  255.30  $&$  141.61 $&$ 4.708830 $&$ -0.313667 $ \\
$71   $&$  395.10  $&$  123.45 $&$ 4.846370 $&$ -0.310110 $ \\
\hline
\hline
\end{tabular}
\end{table}
\begin{table}[ht!]
\centering
\caption{Gate properties, two electrons simulation\label{tab:Gatp_tes}}
\begin{tabular}{lccc}
\hline
\hline
qubit & ECR error                    &  Readout error &  $1$qb gate error  \\
\hline
$60$ & $59\leftrightarrow60 : 0.00419$&$  0.0066        $&$  0.000121$          \\
$59$ & $59\leftrightarrow58 : 0.00625$&$  0.0208        $&$  0.000093  $        \\
$58$ & $71\leftrightarrow58 : 0.00767$&$  0.0225        $&$  0.000119 $         \\
$71$ &                                &$  0.0382        $&$  0.000181   $      \\
\hline
\hline
\end{tabular}
\end{table}

\section{Ground state circuit representation: $N=2$, $\mu=2$}
\label{app:GS_N2mu2}
The Hamiltonian in this sector can be written:
\begin{equation}
\begin{pmatrix}
U\text{-}2\mu & \text{-}t & \text{-}t & 0 \\
\text{-}t & \text{-}2\mu & 0 & \text{-}t \\
\text{-}t & 0 & \text{-}2\mu & \text{-}t \\
0 & \text{-}t & \text{-}t & U\text{-}2\mu
\end{pmatrix},
\label{eq:H}
\end{equation}
where we haved used the basis, \{\ket{1100}, \ket{1001}, \ket{0110}, \ket{0011}\}, and ordered the states as,
$\ket{2\hspace{-0.1cm}\uparrow,2\hspace{-0.1cm}\downarrow,1\hspace{-0.1cm}\uparrow,1\hspace{-0.1cm}\downarrow}$.
The ground state, obtained via diagonalization of (\ref{eq:H}), is:
\begin{equation}
\ket{\Omega} = \frac{1}{\sqrt{2}}\left[\cos\theta\left(\ket{1001}-\ket{0110}\right)+\sin\theta\left(\ket{0011}+\ket{1100}\right)\right],
\label{eq:GS}
\end{equation}
where $\theta = \tan^{-1}(-(\Omega+2\mu)/2t)$, and the ground state energy, $\Omega = 1/2(U-4\mu-\sqrt{U^2+16t^2})$.
For the ease of notation we define, $a\equiv \cos\theta$ and $b\equiv \sin\theta$. 
In addition we introduce the following notation:
\begin{align}
(S_{ij}) =&\frac{1}{2}\left(\mathbb{1}_i \otimes \mathbb{1}_j + X_i \otimes X_j  + Y_i \otimes Y_j + Z_i \otimes Z_j \right)\\
(RY_i(\theta)) =& \cos(\theta/2)\ket{0_i}\bra{0_i} + \sin(\theta/2)\ket{0_i}\bra{1_i} \notag\\ -& \sin(\theta/2)\ket{1_i}\bra{0_i} + \cos(\theta/2)\ket{1_i}\bra{1_i} \\
(C_iC_jX_k) =& \ket{0_i}\bra{0_i}\otimes \mathbb{1}_j \otimes \mathbb{1}_k +  \ket{1_i}\bra{1_i} \otimes (C_jX_k)
\end{align}
$(S_{ij})$ represents the swap operator, which exchanges qubit $i$ with qubit $j$. 
$(RY_i(\theta))$ represents rotation operator about the $y$-axis acting on the $i$th qubit.
$(C_iC_jX_k)$ is the Toffoli gate, where qubits $i$ and $j$ are the control qubits and qubit $k$ is the target qubit. Note that $(\bar{C}_i\bar{C}_jX_k)$ represents the Toffoli gate where the control state is $\ket{0}$ for the control qubits.
 
The states of our basis can be represented using four qubits, which we
order from right to left (i.e. the state $\ket{0110}$ corresponds to $\ket{0_3} \otimes \ket{1_2} \otimes \ket{1_1} \otimes \ket{0_0}$, where the 
subscripts refer to qubit indices). We now aim to represent the ground state as a quantum circuit. Our goal is to obtain
a string of qubit operators acting on the state $\ket{0000}$. This can be achieved by beginning from the ground state
and in a step-by-step manner constructing equivalent states that contain one more qubit operation than the previous
representation, until the state is represented as a string of qubit operations acting on $\ket{0000}$.
Carrying out this procedure we obtain the following circuit representation of $\ket{\Omega}$:
\begin{widetext}
\begin{align}
\ket{\Omega} &= \frac{1}{\sqrt{2}}\left[a(\ket{1001}-\ket{0110}+b(\ket{0011}+\ket{1100}))\right]\notag\\
&=  \frac{1}{\sqrt{2}}(S_{12})\left[a(\ket{1001}-\ket{0110})+b(\ket{0101}+\ket{1010})\right]\notag\\
&=  \frac{1}{\sqrt{2}}(S_{12})(C_1Z_2)\left[a(\ket{1001}+\ket{0110})+b(\ket{0101}+\ket{1010})\right]\notag\\
&=  \frac{1}{\sqrt{2}}(S_{12})(C_1Z_2)(C_1X_2)\left[a(\ket{1001}+\ket{0010})+b(\ket{0101}+\ket{1110})\right]\notag\\
&=  \frac{1}{\sqrt{2}}(S_{12})(C_1Z_2)(C_1X_2)(C_1X_3)\left[a(\ket{1001}+\ket{1010})+b(\ket{0101}+\ket{0110})\right]\notag\\
&=  \frac{1}{\sqrt{2}}(S_{12})(C_1Z_2)(C_1X_2)(C_1X_3)(X_0)\left[a(\ket{1000}+\ket{1011})+b(\ket{0100}+\ket{0111})\right]\notag\\
&=  \frac{1}{\sqrt{2}}(S_{12})(C_1Z_2)(C_1X_2)(C_1X_3)(X_0)(X_3)\left[a(\ket{0000}+\ket{0011})+b(\ket{1100}+\ket{1111})\right]\notag\\
&=  \frac{1}{\sqrt{2}}(S_{12})(C_1Z_2)(C_1X_2)(C_1X_3)(X_0)(X_3)(C_0X_1)\left[a(\ket{0000}+\ket{0001})+b(\ket{1100}+\ket{1101})\right]\notag\\
&=  \frac{1}{\sqrt{2}}(S_{12})(C_1Z_2)(C_1X_2)(C_1X_3)(X_0)(X_3)(C_0X_1)(C_2X_3)\left[a(\ket{0000}+\ket{0001})+b(\ket{0100}+\ket{0101})\right]\notag\\
&=  \frac{1}{\sqrt{2}}(S_{12})(C_1Z_2)(C_1X_2)(C_1X_3)(X_0)(X_3)(C_0X_1)(C_2X_3)(RY_2(2\theta))\left[\ket{0000}+\ket{0001}\right]\notag\\
&=  (S_{12})(C_1Z_2)(C_1X_2)(C_1X_3)(X_0)(X_3)(C_0X_1)(C_2X_3)(RY_2(2\theta))(H_0)\ket{0000}
\end{align}
\end{widetext}
which corresponds to the circuit in Fig.~\ref{fig:2sites_mu2}. We note that this is only one of many possible circuit representations of the ground state, but all such representations are equivalent. 

\section{Ground state circuit representation: $N=4$, $\mu=0$}
\label{app:GS_N4mu0}
For the case of 4 sites, the Hamiltonian is of dimension $16\times16$ (for the sake of brevity we do not write it explicitly here). 
The ground state can be written generically as:
\begin{equation}
\ket{\Omega} = \sum_{i}c_i \ket{\phi}_i.
\label{eq:GS_4site}
\end{equation}
The 16 basis states (using the Qiskit ordering convention, $\ket{4\hspace{-0.1cm}\uparrow,4\hspace{-0.1cm}\downarrow,3\hspace{-0.1cm}\uparrow,3\hspace{-0.1cm}\downarrow2\hspace{-0.1cm}\uparrow,2\hspace{-0.1cm}\downarrow,1\hspace{-0.1cm}\uparrow,1\hspace{-0.1cm}\downarrow}$) and corresponding coefficients are given in Table~\ref{table:GS}.
\begin{table}[ht!]
\caption{Basis states and coefficients of the ground state wavefunction. We include only those elements of the Fock basis
with non-zero coefficients in Eq.~(\ref{eq:GS_4site}). These coefficients are given in the first two columns.}
 \begin{tabular}{r | l | l}
  \hline
  \hline
  $c_i$ && $\ket{\phi}_i$\\
  \hline  
  $c_0$ & $a/2$ &  \ket{10000001}, \ket{00100100} \\ 
  \hline
  -$c_0$ & -$a/2$ & \ket{00011000}, \ket{01000010}\\
  \hline
          $c_1$ & $b/2$ & \ket{00000011}, \ket{00001100}, \ket{00110000}, \ket{11000000}  \\
          \hline
          $c_2$ & $c/\sqrt{8}$& \ket{00001001}, \ket{00100001}, \ket{10000100}, \ket{10010000} \\
          \hline
          -$c_2$ & $c/\sqrt{8}$& \ket{00000110}, \ket{00010010}, \ket{01001000}, \ket{01100000}  \\ 
  \hline\hline
 \end{tabular}
 \label{table:GS}
\end{table}
%
In the same table we also introduce the coefficients $a/2 \equiv c_0 = 0.30581423$, $b/2 \equiv c_1 = 0.14092260$ and $c/\sqrt{8} \equiv c_2 = 0.26136036$. The fact that several elements share the same coefficient is due to symmetries of the ground state wavefunction related to the choice of $\mu = 0$. With these definitions we have the following normalization condition:
\begin{equation}
a^2 + b^2 +c^2 =1,
\end{equation} 
which can be parameterized as follows by introducing the angles $\theta$ and $\phi$,
\begin{equation}
(\cos\phi\sin\theta)^2 + (\sin\phi\sin\theta)^2 +(\cos\theta)^2 =1,
\end{equation} 
 where $\theta\equiv \cos^{-1}(c)$ and $\phi \equiv \cos^{-1}(a/\sin\theta)$.
 
The terms in the ground state can then be rearranged and regrouped to help facilitate the mapping to a quantum circuit:
\begin{widetext}
\begin{align}
\ket{\Omega} 
    &=\frac{1}{2} \ket{0001}_\uparrow \otimes \left[ b\ket{0001}_\downarrow - a\ket{1000}_\downarrow - \frac{c}{\sqrt{2}}\ket{0010}_\downarrow - \frac{c}{\sqrt{2}}\ket{0100}_\downarrow \right] \notag\\
    &+\frac{1}{2}  \ket{0010}_\uparrow \otimes \left[ b\ket{0010}_\downarrow - a\ket{0100}_\downarrow + \frac{c}{\sqrt{2}}\ket{0001}_\downarrow - \frac{c}{\sqrt{2}}\ket{1000}_\downarrow \right]\notag\\
    &+\frac{1}{2}  \ket{0100}_\uparrow \otimes \left[ b\ket{0100}_\downarrow + a\ket{0010}_\downarrow + \frac{c}{\sqrt{2}}\ket{0001}_\downarrow - \frac{c}{\sqrt{2}}\ket{1000}_\downarrow \right]\notag\\
    &+\frac{1}{2}  \ket{1000}_\uparrow \otimes \left[ b\ket{1000}_\downarrow + a\ket{0001}_\downarrow + \frac{c}{\sqrt{2}}\ket{0010}_\downarrow + \frac{c}{\sqrt{2}}\ket{0100}_\downarrow \right],
    \label{eq:GS_4site}
    \end{align}
where we have separated each basis state into a product of spin-$\uparrow$ and spin-$\downarrow$ sectors for clarity (i.e. $\ket{\phi} = \ket{\phi}_\uparrow \otimes \ket{\phi}_\downarrow$). We now aim to rewrite the terms in square brackets in Eq.~(\ref{eq:GS_4site}) using qubit operations to obtain a common term that can be factorized. Taking as an example the first term in square brackets:
\begin{align}
    &\frac{1}{2} \ket{0001}_\uparrow \otimes \left[ b\ket{0001}_\downarrow - a\ket{1000}_\downarrow - \frac{c}{\sqrt{2}}\ket{0010}_\downarrow - \frac{c}{\sqrt{2}}\ket{0100}_\downarrow \right] \notag\\
    &=\frac{1}{2}  (C_4 Z_1)\ket{0001}_\uparrow \otimes\left[ b\ket{0001}_\downarrow - a\ket{1000}_\downarrow + \frac{c}{\sqrt{2}}\ket{0010}_\downarrow - \frac{c}{\sqrt{2}}\ket{0100}_\downarrow \right] \notag \\
    &=\frac{1}{2}  (C_4 Z_1) (C_4 Z_2) \ket{0001}_\uparrow \otimes\left[ b\ket{0001}_\downarrow - a\ket{1000}_\downarrow + \frac{c}{\sqrt{2}}\ket{0010}_\downarrow + \frac{c}{\sqrt{2}}\ket{0100}_\downarrow \right] \notag \\   
    &=\frac{1}{2} (C_4 Z_1)(C_4 Z_2) (C_4 Z_3)\ket{0001}_\uparrow \otimes\left[ b\ket{0001}_\downarrow + a\ket{1000}_\downarrow + \frac{c}{\sqrt{2}}\ket{0010}_\downarrow + \frac{c}{\sqrt{2}}\ket{0100}_\downarrow \right] \notag \\
    &=\frac{1}{2} (C_4 Z_1)(C_4 Z_2)(C_4 Z_3)  (C_4 {S}_{03}) \ket{0001}_\uparrow \otimes\left[ b\ket{1000}_\downarrow + a\ket{0001}_\downarrow + \frac{c}{\sqrt{2}}\ket{0010}_\downarrow + \frac{c}{\sqrt{2}}\ket{0100}_\downarrow \right].   
    \label{eq:t1} 
    \end{align}
The term in brackets in the last line of Eq.~(\ref{eq:t1}) is now identical to the fourth term in brackets of Eq.~(\ref{eq:GS_4site}). Similar manipulations can be done for the other terms in Eq.~(\ref{eq:GS_4site}), and we ultimately obtain:
\begin{align}
   \ket{\Omega}  =&\frac{1}{2} (C_4 Z_1)(C_4 Z_2)(C_4 Z_3)  (C_4 {S}_{03}) \ket{0001}_\uparrow \otimes \left[ b\ket{1000}_\downarrow + a\ket{0001}_\downarrow + \frac{c}{\sqrt{2}}\ket{0010}_\downarrow + \frac{c}{\sqrt{2}}\ket{0100}_\downarrow \right] \notag\\
    &+\frac{1}{2}(C_5 Z_3)(C_5 Z_2)(C_5 {S}_{02}) (C_5 {S}_{13})\ket{0010}_\uparrow \otimes \left[ b\ket{1000}_\downarrow + a\ket{0001}_\downarrow + \frac{c}{\sqrt{2}}\ket{0010}_\downarrow + \frac{c}{\sqrt{2}}\ket{0100}_\downarrow \right]\notag\\
    &+\frac{1}{2}  (C_6 Z_3)(C_6 {S}_{01})(C_6 {S}_{23}) \ket{0100}_\uparrow \otimes \left[ b\ket{1000}_\downarrow + a\ket{0001}_\downarrow + \frac{c}{\sqrt{2}}\ket{0010}_\downarrow + \frac{c}{\sqrt{2}}\ket{0100}_\downarrow \right]\notag\\
    &+\frac{1}{2}  \ket{1000}_\uparrow \otimes \left[ b\ket{1000}_\downarrow + a\ket{0001}_\downarrow + \frac{c}{\sqrt{2}}\ket{0010}_\downarrow + \frac{c}{\sqrt{2}}\ket{0100}_\downarrow \right]
    \end{align}
The use of $C_4, C_5, C_6$ allows us to factorize:
\begin{align}
\ket{\Omega} =&  (C_4 Z_1)(C_4 Z_2)(C_4 Z_3)  (C_4 {S}_{03}) (C_5 Z_3)(C_5 Z_2)(C_5 {S}_{02}) (C_5 {S}_{13})
 (C_6 Z_3)(C_6 {S}_{01})(C_6 {S}_{23}) \notag\\
&\left( \frac{1}{2}\ket{0001}_\uparrow + \frac{1}{2}\ket{0010}_\uparrow  + \frac{1}{2}\ket{0100}_\uparrow  + \frac{1}{2}\ket{1000}_\uparrow  \right) \otimes \left( b\ket{1000}_\downarrow + a\ket{0001}_\downarrow + \frac{c}{\sqrt{2}}\ket{0010}_\downarrow + \frac{c}{\sqrt{2}}\ket{0100}_\downarrow \right)
\end{align}
Now we reduce the first term in parentheses:
\begin{align}
     \frac{1}{2}\ket{0001}_\uparrow + \frac{1}{2}\ket{0010}_\uparrow  + \frac{1}{2}\ket{0100}_\uparrow  + \frac{1}{2}\ket{1000}_\uparrow &= \frac{1}{2}(C_6 X_4)\left( \ket{0001}_\uparrow  + \ket{0010}_\uparrow  + \ket{0101}_\uparrow  + \ket{1000}_\uparrow  \right)\notag\\
     &= \frac{1}{2}(C_6 X_4)(c_6 X_5)\left( \ket{0001}_\uparrow  + \ket{0010}_\uparrow  + \ket{0111}_\uparrow  + \ket{1000}_\uparrow  \right)\notag\\
     &= \frac{1}{2}(C_6 X_4)(c_6 X_5)(\bar C_4 \bar C_5 X_7)\left( \ket{0001}_\uparrow  + \ket{0010}_\uparrow  + \ket{0111}_\uparrow  + \ket{0000}_\uparrow  \right)\notag\\
     &= \frac{1}{2}(C_6 X_4)(c_6 X_5)(\bar C_4 \bar C_5 X_7)(C_4 C_5 X_6)\left( \ket{0001}_\uparrow  + \ket{0010}_\uparrow  + \ket{0011}_\uparrow  + \ket{0000}_\uparrow  \right)\notag\\
     &= (C_6 X_4)(C_6 X_5)(\bar C_4 \bar C_5 X_7)(C_4 C_5 X_6)(H_4)(H_5)\ket{0000}_\uparrow
\end{align}
And the second term:
\begin{align}
     b\ket{1000}_\downarrow + a\ket{0001}_\downarrow + \frac{c}{\sqrt{2}}\ket{0010}_\downarrow + \frac{c}{\sqrt{2}}\ket{0100}_\downarrow
     &= (C_3 X_0) \left( b\ket{1001}_\downarrow + a\ket{0001}_\downarrow + \frac{c}{\sqrt{2}}\ket{0010}_\downarrow + \frac{c}{\sqrt{2}}\ket{0100}_\downarrow\right)\notag\\
     &= (C_3 X_0) (\bar C_0 \bar C_2 X_1) \left( b\ket{1001}_\downarrow + a\ket{0001}_\downarrow + \frac{c}{\sqrt{2}}\ket{0000}_\downarrow + \frac{c}{\sqrt{2}}\ket{0100}_\downarrow\right)\notag\\
     &= (C_3 X_0) (\bar C_0 \bar C_2 X_1) (C_0 RY_3(2\phi))(\bar C_0 H_2)(RY_0(2\theta))\ket{0000}_\downarrow,
\end{align}
where the angles $\phi$ and $\theta$ have been defined above.
Finally we have,
\begin{align}
 \ket{\Omega} =&  (C_4 Z_1)(C_4 Z_2)(C_4 Z_3)  (C_4 {S}_{03}) (C_5 Z_3)(C_5 Z_2)(C_5 {S}_{02}) (C_5 {S}_{13})
     (C_6 Z_3)(C_6 {S}_{01})(C_6 {S}_{23}) \notag\\
     &(C_6 X_4)(C_6 X_5)(\bar C_4 \bar C_5 X_7)(C_4 C_5 X_6)(H_4)(H_5) (C_3 X_0) (\bar C_0 \bar C_2 X_1) (C_0 RY_3(2\phi))(\bar C_0 H_2)(RY_0(2\theta))\ket{00000000}
\end{align}
In order to match the ordering used by Qiskit ($\ket{4\hspace{-0.1cm}\uparrow,4\hspace{-0.1cm}\downarrow,3\hspace{-0.1cm}\uparrow,3\hspace{-0.1cm}\downarrow2\hspace{-0.1cm}\uparrow,2\hspace{-0.1cm}\downarrow,1\hspace{-0.1cm}\uparrow,1\hspace{-0.1cm}\downarrow}$) we renumber the qubit indices to obtain:
\begin{align}
 \ket{\Omega} =&  (C_1 Z_2)(C_1 Z_4)(C_1 Z_6)  (C_1 {S}_{06}) (C_3 Z_6)(C_3 Z_4)(C_3 {S}_{04}) (C_3 {S}_{26})
     (C_5 Z_6)(C_5 {S}_{02})(C_5 {S}_{46}) \notag\\
     &(C_5 X_1)(C_5 X_3)(\bar C_1 \bar C_3 X_7)(C_1 C_3 X_5)(H_1)(H_3) (C_6 X_0) (\bar C_0 \bar C_4 X_2) (C_0 RY_6(2\phi))(\bar C_0 H_4)(RY_0(2\theta))\ket{00000000},
\end{align}
\end{widetext}
which corresponds to the circuit in Fig.~\ref{fig:4sites_mu0}a

\begin{figure}[!ht]
\begin{center}
\includegraphics[width=0.95\columnwidth]{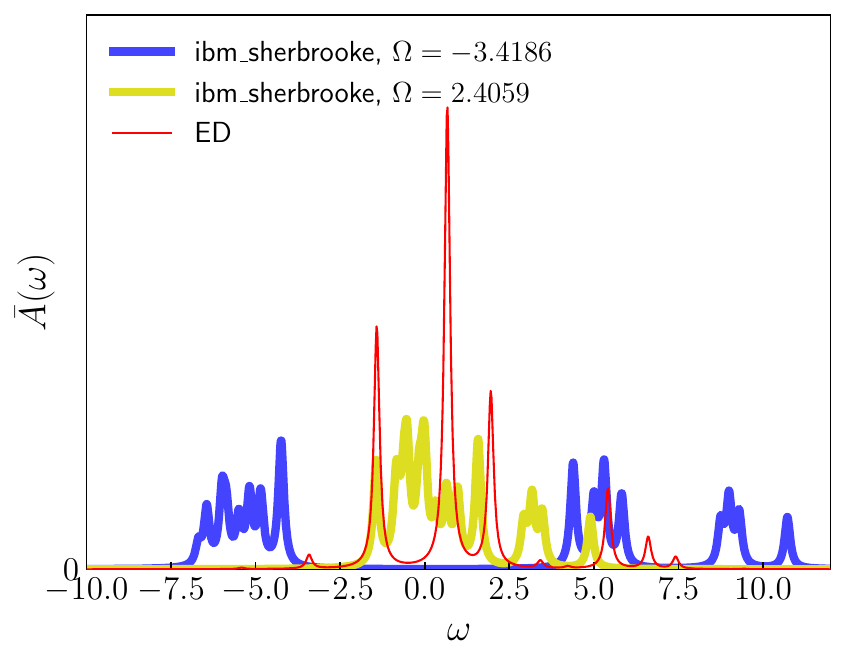}
\end{center}
\caption{Local spectral function with (a) exact ground state energy and (b) ground state energy estimated on quantum device. In this case, all sites are equivalent, therefore we average the results to reduce noise, i.e. $\bar{A}(\omega) = (1/N_s)\sum_i A_{ii}(\omega)$.}
\label{fig:4site_results}
\end{figure}
\section{Results for $N=4$, $\mu=0$ from $\it{ibm}$\_$\it{sherbrooke}$}
\label{app:results_N4mu0}
We present here results for the local spectral function for a $2\times2$ system with 2 electrons. These calculations were carried out on the $\it{ibm}$\_$\it{sherbrooke}$ device, see Fig.~\ref{fig:4site_results}. Already, the ground state energy measured $\Omega^{\it{ibm}\_\it{sherbr.}}_0 = 2.4059$ is quite different than the exact ground state energy $\Omega^\textmd{exact}_0 = -3.4186$. Since $\Omega$ plays an important role (see Eq.~\eqref{eq:Gpm3} for example), we show both results here. Although the quantum simulations predict good results (see Fig.~\ref{fig:4sites_mu0}), the size of the circuit for the ground state is probably too large to carry out the measurement before the quantum state loses its coherence.


%

\end{document}